\documentclass[aps,prc,showpacs,amsmath,amssymb]{revtex4}

\usepackage{epsfig}

\begin{document}


\title{High Tranverse Momentum Suppression in Au+Au Collisions}

\author{D.~E.~Kahana}
\affiliation{31 Pembrook Dr.,
Stony Brook, NY 11790, USA}
\author{S.~H.~Kahana}
\affiliation{Physics Department, Brookhaven National Laboratory\\
   Upton, NY 11973, USA}

\date{\today}  
 
\begin{abstract}
  
Great interest has  attached to recent Au+Au, $\sqrt{s}  = 200$ A
GeV measurements at RHIC,  obtained with the PHENIX, STAR, PHOBOS
and  BRAHMS  detectors.   For  central collisions  and  vanishing
pseudo-rapidity all  experiments indicate a  considerable lowering
in  charged  particle  production  at  mid  to  large  transverse
momenta.   In the  PHENIX  experiment similar  behavior has  been
reported for $\pi^0$ spectra.  In a recent work~\cite{luc4brahms}
on  the presumably  simpler  D+Au interaction,  to be  considered
perhaps as a tune-up for Au+Au, we reported on a hadronic cascade
mechanism  which can  explain the  observed but  somewhat reduced
$p_T$ suppression at higher pseudo rapidity as well as the Cronin
enhancement at mid  rapidity. Here we report on  the extension of
this work to the more massive ion-ion collisions.

\end{abstract}


\maketitle 
\section{Introduction}

The specific question at hand  in this work is the suppression of
medium  to  high  transverse  momentum  yields  observed  in  the
experimental   measurements~\cite{brahms,phenix,phobos,star}  for
Au+Au  at 130 GeV and  200 GeV.   The experimental  results have
focused on the $\eta$ and $p_\perp$-dependence of the ratio

\begin{equation}
\text{R}[\text{AA/NN}] =
\left(\frac{1}{\text{N}_{coll}}\right)
\frac{
[d^2N^{ch}/dp_\perp\,d\eta]\,\,({\text{AA}})
}
{
[d^2N^{ch}/dp_\perp\,d\eta]\,\,({\text{NN}})
}
,
\end{equation}

\noindent  where  $\text{N}_{coll}$  is  a calculated  number  of
binary NN collisions occurring in Au+Au at a designated energy and
centrality. One can  also, of course, just refer  directly to the
data without reference to the ratios.

The simulation code LUCIFER,  developed for high energy heavy-ion
collisions  has  previously been  applied  to  both SPS  energies
$\sqrt{s}=(17.2,20)$ A  GeV~\cite{lucifer1} and to  RHIC energies
$\sqrt{s}=(56,130,200)$  A  GeV ~\cite{lucifer2,luc3}.   Although
nominally intended for dealing  with soft, low $p_T$, interaction
it is possible to introduce  high $p_T$ hadron spectra via the NN
inputs, which form the building blocks of the simulations, and to
then    examine   the    effect   of    rescattering    on   such
spectra~\cite{luc4brahms}.

We  present a  brief description  of the  dynamics of  this Monte
Carlo simulation.  Many other simulations of heavy ion collisions
exist and these are frequently hybrid in nature, using say string
models                in                the               initial
state~\cite{rqmd,rqmd2,bass1,frithjof,capella,werner,ko,ranft}
together with  final state hadronic collisions,  while some codes
are purely partonic~\cite{boal,eskola,wang,wang2,geiger,bass2} in
nature.  Our  approach is closest in  spirit to that  of RQMD and
K.~Gallmeister, C.~Greiner,  and Z.~Xu~\cite{greiner} as  well as
work  by  W.~Cassing~\cite{rqmd,greiner,cassing}.  Certainly  our
results seem to parallel those  of the latter authors.  Both seek
to  separate  initial  perhaps  parton dominated  processes  from
hadronic  interactions  occurring  at  some intermediate  but  not
necessarily late time.

The  purpose of  describing such  high energy  collisions without
introducing the  evident parton nature  of hadrons, at  least for
soft  processes,  was  to  set  a baseline  for  judging  whether
deviations  from the simulation  measured in  experiments existed
and  could  then  signal  interesting  phenomena.   The  division
between soft  and hard processes,  the latter being  in principle
described  by  perturbative  QCD,  is  not  necessarily  easy  to
identify in  heavy ion data,  although many authors  believe they
have    accomplished     this    within    a    gluon-saturation
configuration~\cite{saturation,cgc1,cgc2}.   For  both D+Au  and
Au+Au systems we separate the  effects of a second stage, a lower
energy  hadronic  cascade,  from  those  of the  first  stage,  a
parallel   rather   than    sequential   treatment   of   initial
(target)-(projectile) NN interactions.

In the  present work, even absent some  energy-loss effects which
one might anticipate, we  do find considerable suppression of the
Au+Au transverse momentum spectrum  at cental rapidity. One might
say  that  the  first  stage  of our  simulation,  involving  the
parallel interaction of  the initially present nucleons, produces
a ``hot-gas''  of prehadrons which are considerably  cooled in an
inevitable  final state  cascade.  This  cooling  constitutes the
observed  ``jet'' suppression,  a  suppression, not  surprisingly
appreciably       greater      for      Au+Au       than      for
D+Au~/cite{brahms2,luc4brahms}.  The  second  stage  II,  a  true
cascade also, critically includes energy loss effects. The latter
inevitably soften  at least the  lower $p_\perp$, as  II involves
considerably reduced energy processes.

\section{The Simulation}
\subsection{Stage I}

The first  stage I of LUCIFER considers  the initial interactions
between the separate  nucleons in the colliding ions  A+B, but is
not a cascade.  The  totality of events involving each projectile
particle  happen  essentially  together   or  one  might  say  in
parallel.   Neither  energy   loss  nor  creation  of  transverse
momentum  ($p_\perp$)  are  permitted  in  stage  I,  clearly  an
approximation.           A          model          of          NN
collisions~\cite{lucifer1,lucifer2},   incorporating  most  known
inclusive cross-section and multiplicity data, guides stage I and
sets up the initial conditions for stage II.  The two body model,
clearly an  input to  our simulation, is  fitted to  the elastic,
single   diffractive  (SD)   and  non-single   diffractive  (NSD)
aspects~\cite{goulianos}   of   high   energy   $PP$   collisions
~\cite{ua5,ua1}  and  $P\bar   P$  data~\cite{fermilab}.   It  is
precisely  the  energy   dependence  of  the  cross-sections  and
multiplicities  of  the  NN  input  that led  to  our  successful
prediction~\cite{lucifer2,phobos1}  of the rather  small ($13\%$)
increase   in    $dN^{ch}/d\eta$   between   $\sqrt{s}=130$   and
$\sqrt{s}=200$ A GeV, seen in the PHOBOS data~\cite{phobos3}. 

We  find that  the  addition  to stage  I  baryons of  transverse
momentum by  collision dependent random  walk produces in  fact a
rather  ``hot'' gas  of mesons,  in effect  a  strong A-dependent
Cronin~\cite{cronin} effect,  a gas which  is subsequently cooled
by  the  final  state  cascading  collisions in  stage  II.   The
comparison  of the initial  and final  $p_T$ spectra  provides an
alternative   measure   of  suppression   to   the  above   ratio
{R}[{AA/NN].

A history of  the collisions that occur between  nucleons as they
move along straight  lines in stage I is  recorded and later used
to guide determination  of multiplicity.  Collision driven random
walk  in $p_\perp$  fixes the  $p_\perp$  to be  ascribed to  the
baryons  at the  start of  stage II.   The  overall multiplicity,
however, is  subject to a  modification, based, as we  believe on
natural physical requirements~\cite{lucifer2}.

If a sufficiently hard  process, for example Drell-Yan production
of  a lepton  pair at  large mass  occurred, it  would lead  to a
prompt  energy loss  in stage  I.  Hard  quarks and  gluons could
similarly be  entered into the particle lists  and their parallel
progression followed.  This has not yet been done.  One viewpoint
and justification for our approach is to say we attempt to ignore
the direct effect  of colour on the dynamics,  projecting out all
states  of the  combined  system possessing  colour.   In such  a
situation  there  should  be  a duality  between  quark-gluon  or
hadronic treatments. 
 
The  collective/parallel method  of treating  many  NN collisions
between the target and projectile is achieved by defining a group
structure for  interacting baryons.  This is  best illustrated by
considering a prototype  proton-nucleus (P+A) collision.  A group
is  defined  by spatial  contiguity.   A  proton  at some  impact
parameter  $b(\bar{x}_\perp)$  is  imagined  to  collide  with  a
corresponding  `row'  of   nucleons  sufficiently  close  in  the
transverse  direction to the  straight line  path of  the proton,
{\it   i.~e.}~within   a  distance   corresponding   to  the   NN
cross-section.   In   a  nucleus-nucleus  (A+B)   collision  this
procedure is generalized by making  two passes: on the first pass
one includes all  nucleons from the target which  come within the
given  transverse distance  of some  initial  projectile nucleon,
then on the  second pass one includes for  each target nucleon so
chosen, all of those  nucleons from the projectile approaching it
within the  same transverse distance.  This  totality of mutually
colliding   nucleons,   at   more   or  less   equal   transverse
displacements,  constitute  a  group.  The  procedure  partitions
target and projectile nucleons into a set of disjoint interacting
groups as well as a set of non-interacting spectators in a manner
depending on the overall  geometry of the A+B collision.  Clearly
the largest groups in P+A will,  in this way, be formed for small
impact parameters  $b$; while for the  most peripheral collisions
the groups  will almost always  consist of only one  colliding NN
pair. Similar conclusions hold in the case of A+B collisions.
     
In stage II of the  cascade we treat the entities which rescatter
as  prehadrons.  These  prehadrons, both  baryonic or  mesonic in
type, are not the  physical hadron resonances or stable particles
appearing in  the particle  data tables, which  materialize after
hadronisation.   Importantly prehadrons  are allowed  to interact
starting   at    early   times,   after    a   short   production
time~\cite{boris}, nominally  the target-projectile crossing time
$T_{AB} \sim R_{AB}/\gamma$.  The mesonic prehadrons are imagined
to have  ($q \bar  q$) quark content  and their  interactions are
akin to  the dipole interactions included in  models relying more
closely  on explicit  QCD~\cite{boris,mueller},  but are  treated
here  as colourless  objects.   

Some  theoretical  evidence   for  the  existence  of  comparable
colourless structures is  given by Shuryak and Zahed~\cite{zahed}
and  by certain lattice  gauge studies~\cite{lattice}.   In these
latter  works  a basis  is  established  for  the persistence  of
loosely  bound  or  resonant   hadrons  above  the  QCD  critical
temperature $T_c$ to $T  \sim (1.5-2.0)\times T_c$.  This implies
a persistence to much higher transverse energy densities $\rho(E)
\sim (1.5-2.0)^4  \rho_c$, hence  to the early  stages of  a RHIC
collision.  Accordingly  we have incorporated into  stage II {\it
hadron  sized}  cross-sections  for  the  interactions  of  these
prehadrons,  although early  on it  may  in fact  be difficult  to
distinguish  their colour  content.   Such larger  cross-sections
indeed  appear  to  be  necessary  for  the  explanation  of  the
apparently   large    elliptical   flow   parameter    found   in
measurements~\cite {molnar,flow}.

The  prehadrons, which when  mesonic may  consist of  a spatially
close, loosely correlated quark  and anti-quark pair, are given a
mass spectrum  between $m_\pi$ and $1$  GeV, with correspondingly
higher upper  and lower  limits allowed for  prehadrons including
strange  quarks.  The  Monte-Carlo  selection of  masses is  then
governed by a Gaussian distribution,
\begin{equation}
P(m)= \exp(-(m-m_0)^2/w^2),
\end{equation}
with $m_0$ a selected  center for the prehadron mass distribution
and $w=m_0/4$  the width.  The non-strange  mesonic prehadrons is
taken at  $m_0 \sim  500$ MeV, and  for strange at  $m_0\sim 650$
MeV. Small changes in $m_0$  and $w$ have little effect since the
code is constrained to fit hadron-hadron,  data.

Too high an  upper limit for $m_0$ would  destroy the soft nature
expected for most prehadron  interactions when they finally decay
into  `stable'  mesons.   The   same  proviso  is  in  place  for
prebaryons which are restricted to  a mass spectrum from $m_N$ to
$2$ GeV.  However, in the present calculations the prebaryons are
for simplicity taken just to  be the normal baryons.  The mesonic
prehadrons  have  isospin   structure  corresponding  to  $\rho$,
$\omega$, or $K^*$, while the  baryons range across the octet and
decuplet.

Creating  these intermediate  degrees of  freedom at  the  end of
stage I  simply allows the original nucleons  to distribute their
initial energy-momentum  across a larger basis of  states or Fock
space, just  as is done in  string models, or for  that matter in
partonic   cascade   models.    Eventually,  of   course,   these
intermediate  objects decay  into physical  hadrons and  for that
purpose we assign a  uniform decay width $\sim \Gamma_{f}$, which
then plays the  role of a hadronisation or  formation time, $\sim
1/(\Gamma_{f})$.

\subsection{Elementary Hadron-Hadron Model}

The underlying  NN interaction structure involved in  I has been
introduced   in   a  fashion   dictated   by  the   proton-proton
modeling~\cite{goulianos}.   A  division  is made  into  elastic,
single    diffractive    (SD)    and    non-single    diffractive
components. Fits  are obtained  to the existing  two-nucleon data
over  a  broad  range  of  energies  (sqrt(s)),  using  the  same
prehadrons introduced above. No  rescattering, only decay of these
intermediate   structures   is  permitted   in   the  purely   NN
calculation.   Specifically   the  meson-meson  interactions  are
scaled  to $4/9$  of the  known  NN cross-sections,  thus no  new
parameters  are  invoked.  Indeed,  since  only  known data  then
constrains  the  prehadronic  interaction,  this  approach  is  a
parameter-free input to the AA dynamics.

\subsection{Groups}

Energy  loss  and  multiplicity  in  each group  of  nucleons  is
estimated from  the straight line collision  history.  To repeat,
transverse momentum  of prebaryons is  assigned by a  random walk
having  a number  of  steps  equal to  the  number of  collisions
suffered.   The  multiplicity  of  mesonic prehadrons  cannot  be
similarly directly estimated from  the number of NN collisions in
a group.   We argue~\cite{gottfried} that  only spatial densities
of generic prehadrons~\cite{lucifer1,lucifer2} below some maximum
are  allowable,  {\it  viz.}   the prehadrons  must  not  overlap
spatially at the  beginning of stage II of  the cascade.  The KNO
scaled multiplicity distributions, present in our NN modeling are
sufficiently  long-tailed  that imposing  such  a restriction  on
overall multiplicity can for larger nuclei affect results even in
P+A  or D+A  systems.  In  earlier  work~\cite{lucifer1,luc3} the
centrality dependence of $dN/d\eta$ distributions for RHIC energy
Au+Au  collisions   was  well  described  with   such  a  density
limitation  on  the prehadrons,  which  was  not  carried out  as
meticulously as  in the present work, especially  with to respect
to highly peripheral collision.

Importantly,  the cross-sections  in prehadronic  collisions were
assumed to be the same size as hadronic, {\it e.~g.}~meson-baryon
or meson-meson  {\it etc.}~, at  the same center of  mass energy,
thus introducing  no additional  free parameters into  the model.
Where the  latter cross-sections or their  energy dependences are
inadequately known we  employed straightforward quark counting to
estimate the scale.  In both SPS, Pb+Pb and  RHIC Au+Au events at
several energies it was sufficient to impose this constraint at a
single energy.  The inherent  energy dependence in the KNO-scaled
multiplicities of the NN inputs and the geometry then take over.

\subsection{High Transverse Momenta}

One  question which  has yet  to be  addressed concerns  the high
$p_\perp$  tails  included in  our  calculations.  In  principle,
LUCIFER  is  applicable  to  soft processes  {\it  i.~e.}~at  low
transverse momentum.  Where the cutoff in $p_\perp$ occurs is not
readily apparent. In any case we can include high $p_\perp$ meson
events through  inclusion in the  basic hadron-hadron interaction
which  is  of  course  an  input  rather than  a  result  of  our
simulation.   Thus   in  Fig(1)  we  display   the  NSD  $(1/2\pi
p_\perp)(d^2N^{charged}/dp_\perp\,d\eta)$   from  UA1~\cite{ua1}.
One can use  a single exponential together with  a power-law tail
in $p_\perp$, or alternatively two exponentials, to achieve a fit
of the  output in PP to  UA1 $\sqrt{s}$=200 GeV  data. A sampling
function of the form
\begin{equation}
f = p_\perp (a \exp(-p_\perp/w) + b / ((1 + (r / \alpha)^ \beta)),    
\end{equation}
gives a  satisfactory fit to the PP data in the Monte-Carlo.

Additionally,   since  we  constrain   our  comparisons   to  the
production of neutral pions in  Au+Au we also present, in Fig.(2)
the  PHENIX~/cite{phenixpp}   midrapidity  $p_t$  yield   for  NN
together  with  our representation  of  this  spectrum.  These  NN
generated $p_\perp$  spectra, inserted into the  code, were first
applied    to    the     meson    $p_\perp$    distribution    in
D+Au~\cite{luc4brahms} and now of course to Au+Au.  No correction
is  made for  possible  energy  loss in  stage  I, an  assumption
parallel  to  that  made  by   the  BRAHMS  and  all  other  RHIC
experiments,  in analysing  $p_\perp$ spectra  and multiplicities
irrespective of low or high values.  However some explicit energy
loss is  present in the collisions  of stage II,  in the dynamics
through  energy conservation,  but still  employing the  NN meson
representation introduced above.

Since  we impose  energy-momentum conservation  in each  group, a
high $p_\perp$  particle having say, several  GeV/c of transverse
momentum,  must   be  accompanied  in   the  opposite  transverse
direction   by  one   or  several   compensating   mesons.   Such
high-$p_\perp$  leading particles  are not  exactly jets,  to the
extent that  they did not  originate in our simulation  from hard
parton-parton  collisions,  but  they  yield  much  of  the  same
observable experimental behaviour. It must be emphasized that the
totality  of $p_\perp$  events  is small,  certainly  for the  NN
collisions seen in Figs.(1,2)) and also as we will see for any AA
events. In fact  some $90\%$ of all final  given $p_\perp$ yields
occurs for $p_\perp \le 0.7$.  This implies that our treatment of
such processes  is indeed a  perturbation, unlikely to  alter the
overall dynamics.

\subsection{Initial Conditions for II}

The final operation  in stage I is to  set the initial conditions
for the hadronic cascade  in stage II.  The energy-momentum taken
from the initial baryons and shared among the produced prehadrons
is  established  and an  upper  limit  placed  on the  production
multiplicity   of  prehadrons  and   normal  hadrons.    A  final
accounting of  energy sharing is  carried out through  an overall
4-momentum conservation  requirement.  We emphasize  that this is
carried out separately within each group of interacting nucleons.

The spatial  positioning of the  particles at this time  could be
accomplished in a  variety of ways.  We have  chosen to place the
prehadrons in each group  inside a cylinder, initially having the
longitudinal size of the nucleus, for a A+B collision, and having
the initial  longitudinal size of the interaction  region at, and
then  allowing the  cylinder to  evolve freely  according  to the
longitudinal momentum distributions, for a fixed time $\tau_{f}$,
defined in the rest frame of  each group.  At the end of $\tau_f$
the multiplicity of  the prehadrons is limited so  that, if given
normal hadronic  sizes $\sim (4\pi/3) (0.7)^3$  fm$^{3}$, they do
not overlap  within the cylinder.Such  a limitation in  density is
consonant with the general  notion that produced hadrons can only
exist when  separated from the  interaction region in  which they
are generated.

Up  to this  point  longitudinal boost  invariance is  completely
preserved,  since stage  I  is carried  out  using straight  line
paths.  The technique of defining the evolution time in the group
rest frame  is essential to minimizing  residual frame dependence
which inevitably arises in any cascade, hadronic or partonic,when
transverse momentum is  considered due to the finite  size of the
colliding   objects  implied   by   their  non-zero   interaction
cross-sections.
       
\section{Stage II: Final State Cascade}

Stage  II is  as stated  a straightforward  cascade in  which the
prehadronic  resonances  interact  and  decay as  do  any  normal
hadrons  present or  produced during  this  cascade.  Appreciable
energy having being finally transferred to the produced particles
these  `final  state' interactions  occur  at considerably  lower
energy than the initial nucleon-nucleon collisions of stage I. As
pointed out,  during stage II  the interaction and decay  of both
prehadrons  and hadrons  is allowed.   In the  present  case, for
Au+Au,  the effect  of prehadron-prehadron  interaction  is truly
appreciable.

We are  then in a position  to present results for  Au+Au 200 GeV
collisions.   These  appear  in  Figs(3--6)  for  various  double
differential  tranverse  momenta   spectra  or  their  derivative
ratios.   Most  contain  comparisons with  PHENIX~\cite{phenixau}
$\pi^0$ measurements.   In future  work we consider  also charged
data, but there proton spectra play an increasingly larger role.

The initial  conditions created to start the  final cascade could
have perhaps  been arrived  at through more  traditional, perhaps
partonic, means.  The second stage would then still proceed as it
does here.  We reiterate that  our purpose has been to understand
to what extent the results  seen in Figures (1-6) are affected by
stages  I and  II  separately.  {\it  i.~e.}~do  they arise  from
initial or from final state interactions.

\section{Results: Comparison with Data}

Fig(3) contains the simulated $\pi^0$ transverse momenta spectrum
for Au+Au at  $\eta=0$ alongside the PHENIX data~cite{phenixpau}.
To  a large  extent  the suppression  observed experimentally  is
paralleled  by the simulated  calculations.  The  production time
$\tau_p$ introduced above is given two values $2(2R_{Au}/\Gamma)$
and twice this value,  indicating the variation with this initial
state time, a parameter in  our modeling.  The $\Gamma$'s are the
longitudinal  Lorentz factors  defined above  and  introduced for
each  baryon  group  separately  in  its  rest  frame.   For  the
symmetric Au+Au  collision this distinction  by group is  only of
small consequence.

What conclusions are to be drawn from these first results? Clearly
one  cannot  ignore final  state  cascading.   Moreover, for  the
assumptions  we have  made, the  most crucial  being  the perhaps
early commencement  of such cascading, the  suppression cannot be
considered as necessarily a  sign for production of a quark-gluon
plasma:  perhaps only  a  prehadron dominated  medium after  some
initial  delay.   Since  the   PQCD  approach  is   clearly  more
fundamental, provided a clear  treatment of soft processes can be
included, one  cannot rule out  its basis of  interpretation. But
surely it  is interesting to  pursue an alternative,  albeit more
phenomenological  nuclear-system-oriented  view.   A view  which
simply  suggests that  more detailed  and definite  signs  of QCD
plasma must perhaps be sought.

It is instructive to  deconstruct the elements of the simulation,
~i.e. to separate  the spectrum at the conclusion  of I from that
resulting  from  both I+II.   In  Fig.(4)  the $\pi0$  transverse
momentum  yield  is  shown   for  both  these  cases  against  the
experimental  data.   It is  immediately  evident  that the  many
virtual  NN  collisions  in  stage  I  produce  a  much  elevated
$p_\perp$ output and that this is in turn reduced by more than an
order   of  magnitude  by   collisions,  with   presumably  other
prehadrons, in  II. Part of  this effect is through  inclusion in
the  dynamics   of  at  least  a kinematical   treatment  of  energy
loss.  Thus above we  referred to  an initial  hot gas  cooled by
expansion and final state interactions.

One might well turn this  around and declare that the final state
scattering of  a given prehadron  with comovers has cut  down the
Cronin effect,  a reduction  which suggests the  applicability of
the term `jet suppression' by final state interactions. One notes
parenthetically  that  particles   lost  at  high  $p_\perp$  are
compensated for by an increase at the lowest $p_\perp$'s.

A second  and equally important  criterion for the  simulation is
the   maximum  densities  created   as  initial   conditions  for
II. Fig(5)  casts some  light on this  and on another  issue, the
actual  transverse  energy  density  attributed to  the  earliest
stages  of the  collision. We  have included  in this  figure the
charged $dN/d\eta$  spectra: (a)  for the totality  of ``stable''
mesons in I+II, (b) the same  result for I alone when only decays
of prehadrons  are permitted, and finally (c)  for the prehadrons
in I  only with no decays. It  is evident that some  $2/3$ of the
summed  tranverse energy  $E_\perp$  is generated  in the  second
expanding   phase  II   when  the   system  is   increasing  both
longitudinally and  transversely.  The initial  $E_\perp$ is then
reduced commensurably, falls well  below the Bjorken limit and is
hence not  all available  for initial ``plasma''  generation.  In
present calculations at the  initialisation of II, and keeping in
mind the average masses assigned to prehadrons, $0.5 to 0.6$ GeV,
the ambient transverse energy  densities are $\le 1.8$ GeV/$fm^3$)
for the shortest initial time $\tau_p$ chosen and correspondingly
less for longer times.

It is also clear that the density of prehadrons, each of which in
I as seen  from Fig.(5) decays into some  2.5 stable hadrons. The
rather  lengthy formation  time assigned,  $\tau_f \sim  1$ fm/c,
ensures that  these stable particles enter only  sparingly in the
dynamics of II. Thus the  bugbear of too much rescattering is
reduced to manageable proportion.


\section{Conclusions}

It is hard  to conclude definitively from what  is presented here
that the  standard pQCD  + some soft  process treatment is  not a
more fundamental approach. But  at question is just this coupling
to soft processes, not  unrelated to possible early appearance of
an excitation spectrum  of hadronic-like structures.  The latter,
if  generated  sufficiently early  may  alter  the  role of  soft
processes even  on high $p_\perp$ objects passing  through a much
more dense cloud of soft hadronic structures.  Certainly there is
still a silent elephant  lurking in the dynamics, the observation
of    rather    large    elliptical    flow    in    the    meson
spectrum~\cite{flow}. These flow measurements are most easily 
achieved theoretically if hadronic coss-sections are present.  

In  the  work  on  D+Au~\cite{luc4brahms}  the  use  of  such  an
excitation  spectrum  exposed most  clearly  the  simple role  of
geometry  in ratio of  BRAHMS $p_\perp$  spectra.  For  Au+Au its
presence  may  muddy the  waters,  but  surely more  experimental
exploration  is  required.  Certainly  the  RHIC experiments  are
probing  unusual  nuclear matter,  at  high  hadronic and  energy
density, and in exciting terms.

It  would  seem  however  that  the  direct  attempt  at  a  pQCD
explanation of this behaviour must claim that, at the very least,
all soft mesons are produced in essentially hard collisions.  The
presentation here provides an interesting case for relying on the
geometry of soft,  low $p_\perp$, processes, essentially mirrored
in hard processes, to produce  the major features of the D+Au and
Au+Au   data.   True   enough,  the   high  $p_\perp$   tails  in
distributions  are   merely  tacked  on  in   our  approach,  but
legitimately  so   by  using  the   NN  data  as  input   to  the
nucleus-nucleus cascade.  In any case one should  again very much
be  cognizant of  the small  number of  high  $p_\perp$ particles
present in  even cental collisions. Some $5\%$  of the integrated
spectrum of mesons comes from $p_\perp \ge 1$ GeV.

\section{Acknowledgements}
This  manuscript  has  been  authored  under  the  US  DOE  grant
NO. DE-AC02-98CH10886. One of  the authors (SHK) is also grateful
to the  Alexander von Humboldt Foundation, Bonn,  Germany and the
Max-Planck   Institute  for   Nuclear  Physics,   Heidelberg  for
continued  support and hospitality.   Useful discussion  with the
BRAHMS,  PHENIX, PHOBOS  and STAR  collaborations  are gratefully
acknowledged,  especially with  C.~Chasman, R.~Debbe,F.~Videbaek.
M.~T.~Tannenbaum, T.~Ulrich and J.~Dunlop.

\vfill\eject

\begin{figure}
\vbox{\hbox to\hsize{\hfil
\epsfxsize=6.1truein\epsffile[0 0 561 751]{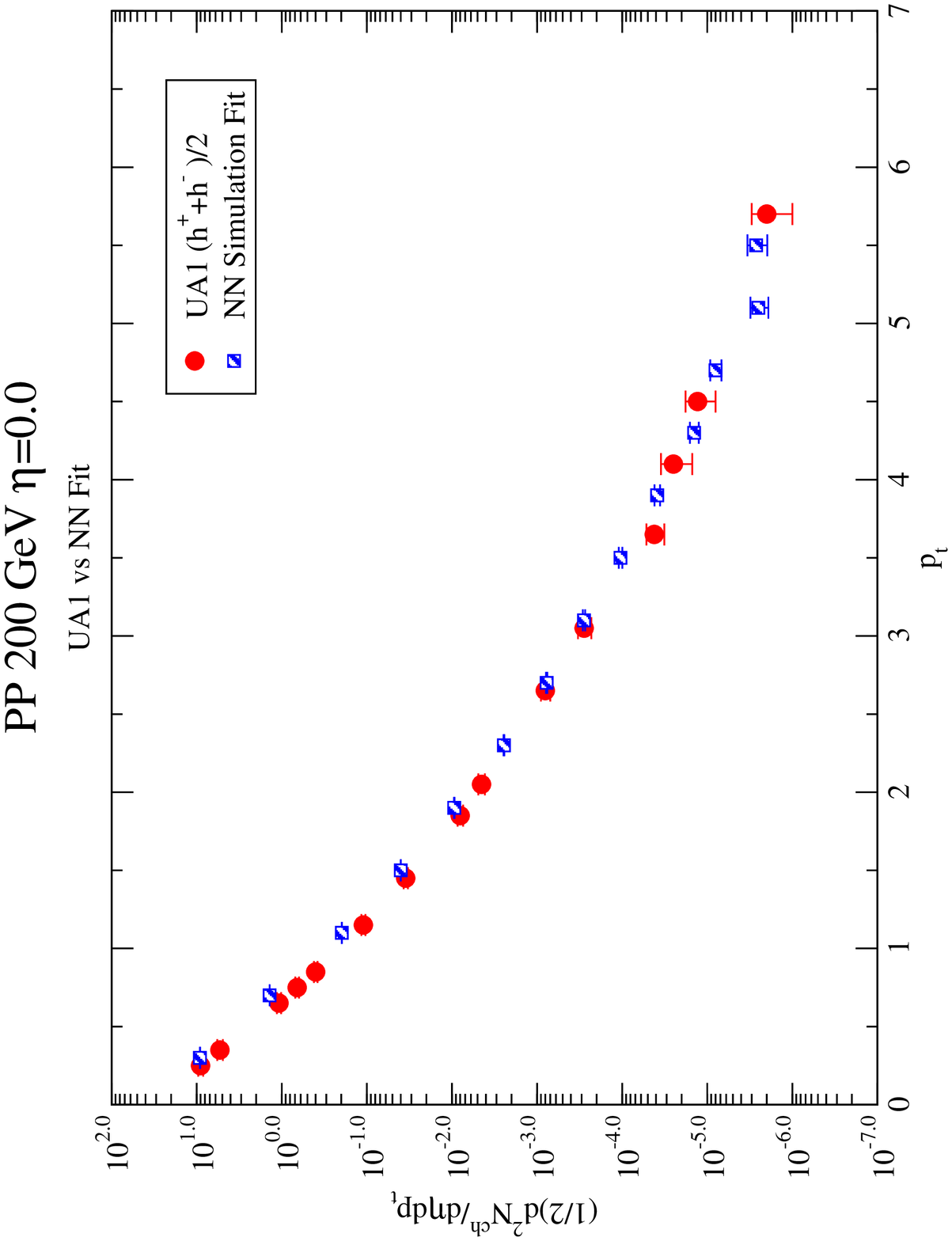}
\hfil}}
\caption[]{PP Pseudorapidity  spectra: Comparison of  UA1 minimum
bias  200 GeV  NSD  data~/cite{ua1} with  an appropriate  LUCIFER
simulation. The latter is properly a fit to the experiment and an
input to  the ensuing AA  collisions; thus does not  constitute a
`set' of free parameters.}
\label{fig:Fig.(1)}
\end{figure}
\clearpage

\begin{figure}
\vbox{\hbox to\hsize{\hfil
\epsfxsize=6.1truein\epsffile[0 0 561 751]{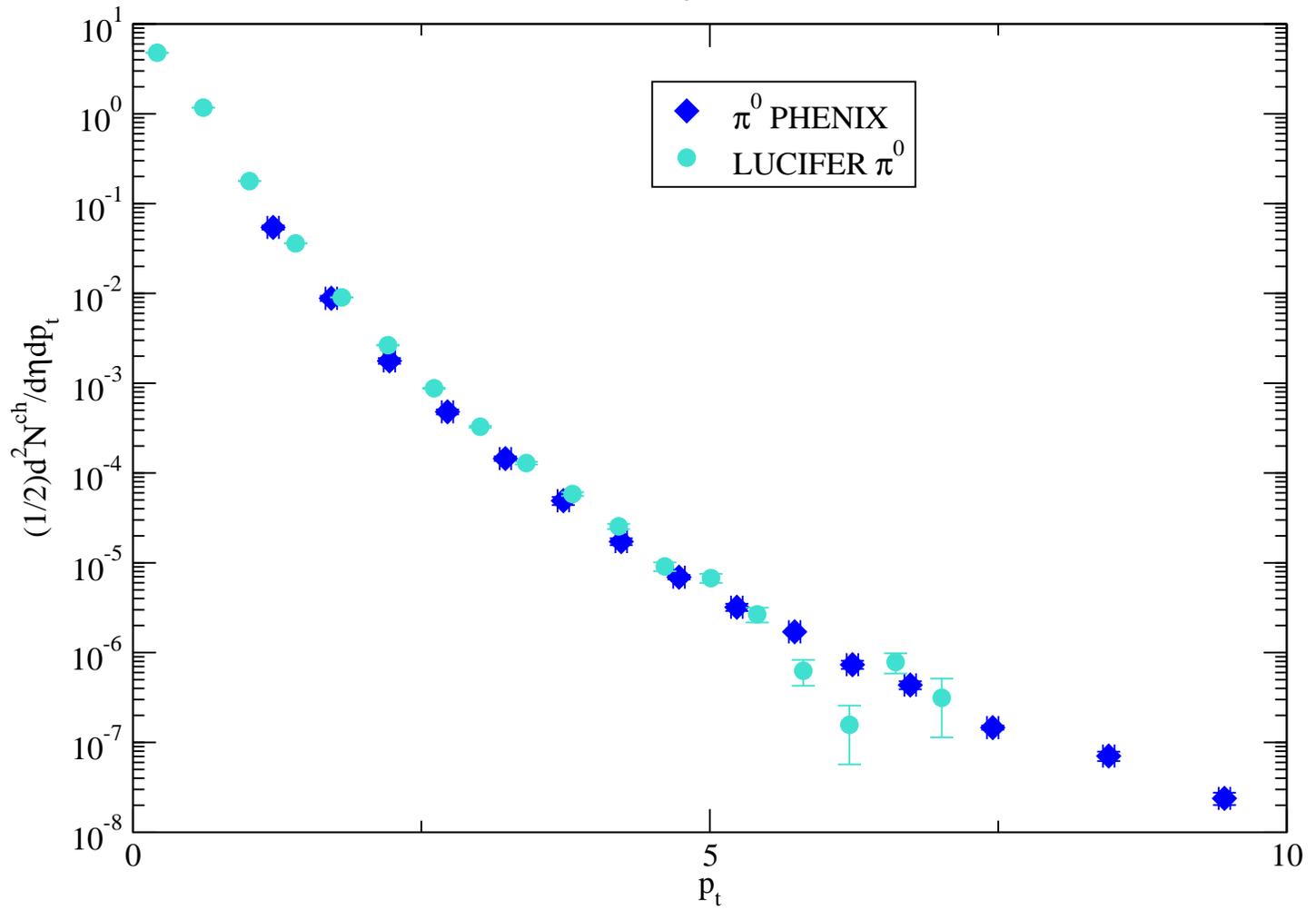}
\hfil}}
\caption[]{A  similar transverse  momentum  $\pi^0$ spectrum  from
PHENIX PP~\cite{phenixpp} vs simulation.}
\label{fig:Fig.(2)}
\end{figure}
\clearpage

\begin{figure}
\vbox{\hbox to\hsize{\hfil
\epsfxsize=6.1truein\epsffile[0 0 561 751]{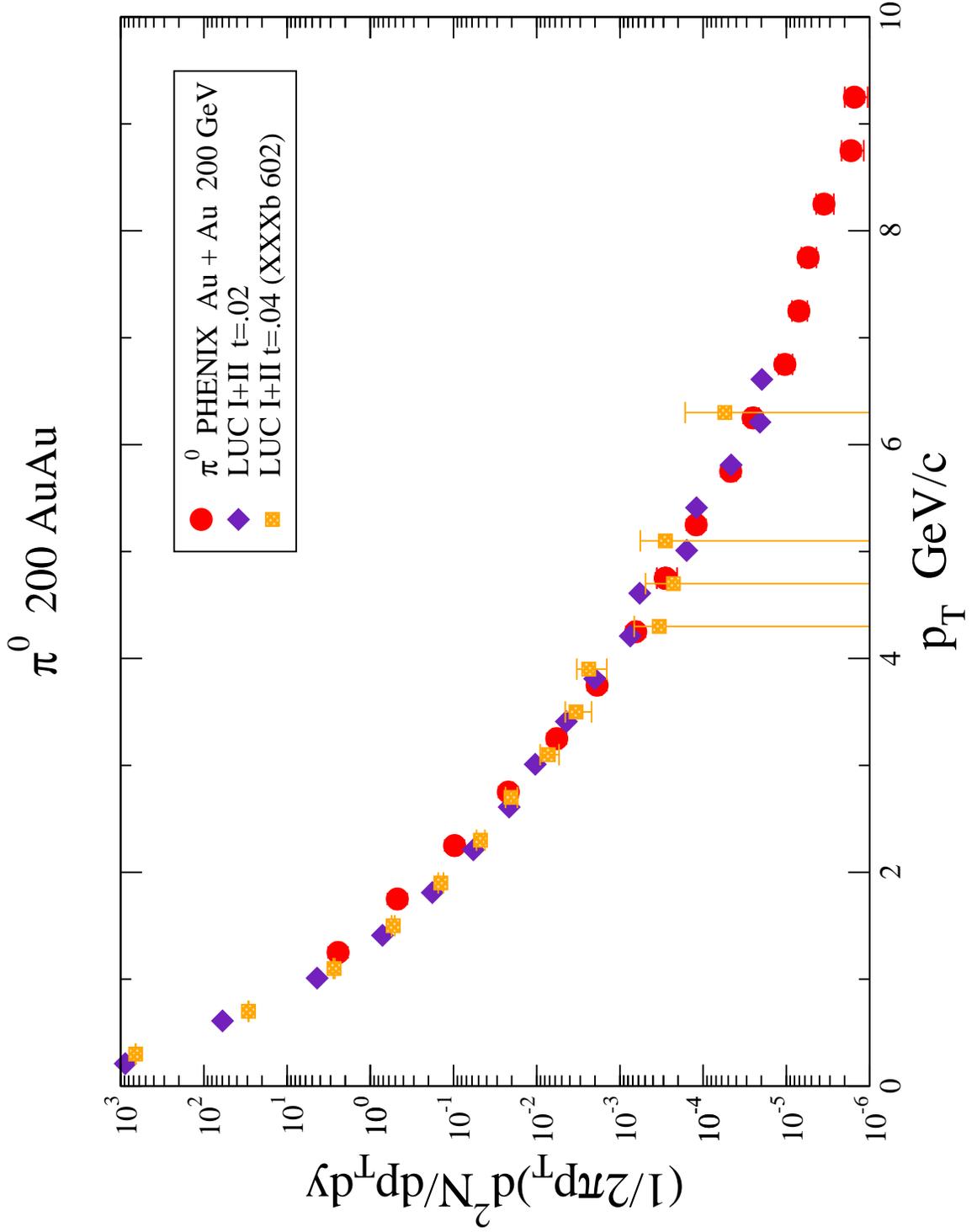}
\hfil}}
\caption[]{Central PHENIX $\pi^0$ 200 GeV for Au+Au vs simulation
for   two   choices  of   $\tau_p$,   the  prehadron   production
time. Centrality  for PHENIX  is here $0\%-10\%$,  roughly impact
parameter $b=4.25$ fm. for the simulation.}
\label{fig:Fig.(3)}
\end{figure}
\clearpage

\begin{figure}
\vbox{\hbox to\hsize{\hfil
\epsfxsize=6.1truein\epsffile[0 0 561 751]{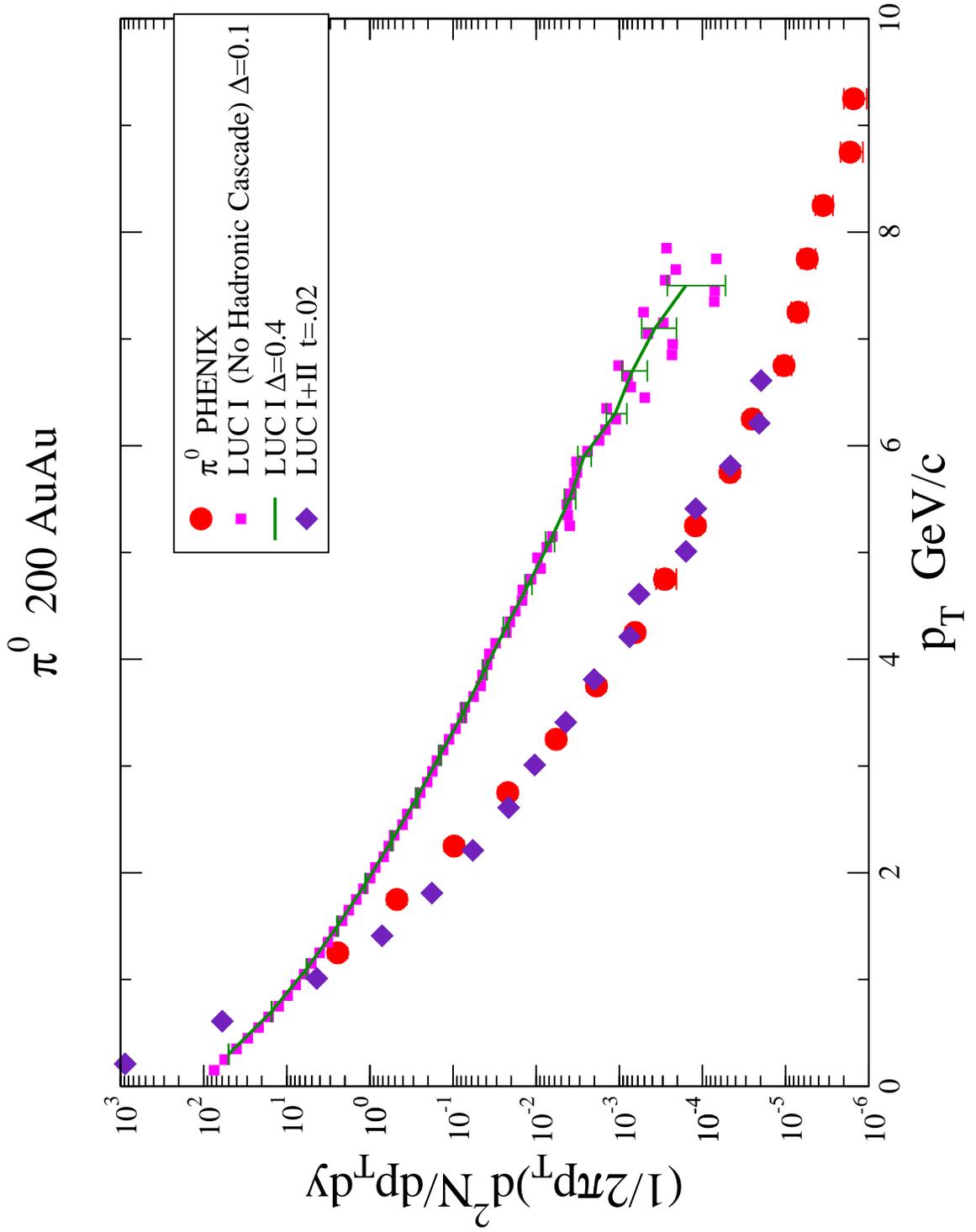}
\hfil}}
\caption[]{The $\pi0$  transverse momenta yields for  stage I, no
final  cascade, vs those  for the  full  stage I+II  calculation.
Clearly there  is considerable suppression in  the final cascade.
Recalling that the experimentalists quote a `direct' suppression
of $5-6$ for  the ratio in Eqn.(1),  there is at the end  of I an
enhancement $\sim 2.5-3$, {\it i.~e.} a Cronin effect in this stage.}
\label{fig:Fig.(4)}
\end{figure}
\clearpage

\begin{figure}
\vbox{\hbox to\hsize{\hfil
\epsfxsize=6.1truein\epsffile[0 0 561 751]{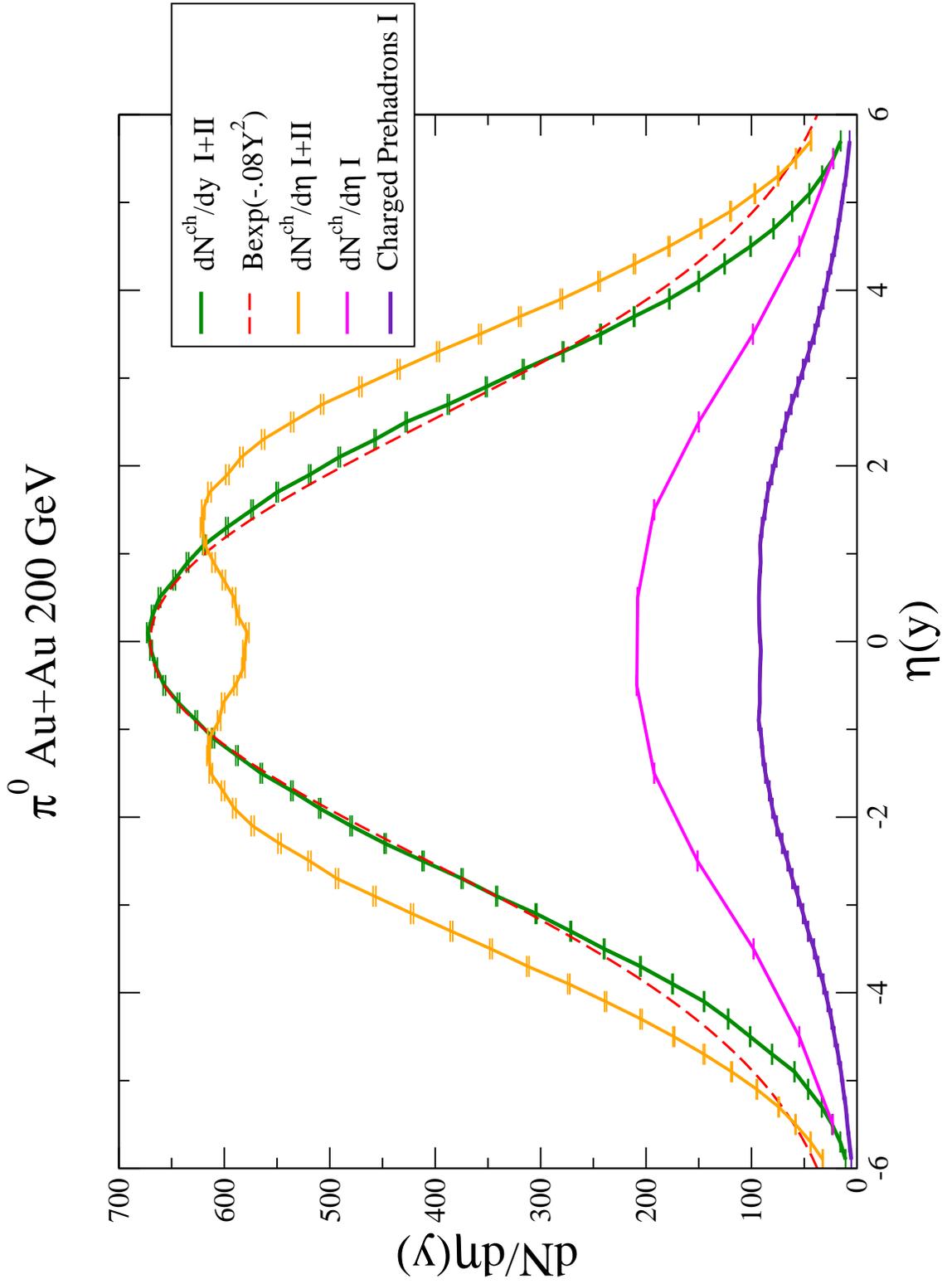}
\hfil}}
\caption[]{Pseudorapidity  and  rapidity   spectra  for  charged  mesons  and
prehadrons  at various  stages  of the  collision.  The gaussian  fit to  the
charged     pion    rapidity     distribution     approximates    preliminary
BRAHMS~\cite{brahmsfwhmprelim}  results, at least in its FWHM.}

\label{fig:Fig.(5)}
\end{figure}
\clearpage
\end{document}